\documentclass{article}
\usepackage{amsmath,amsfonts,amsthm}
\usepackage{graphics,graphicx}

\newcommand{\R}{\mathbb R}

\newcommand{\E}{\mathbb E}
\newcommand{\Ebb}{\mathbb E}
\newcommand{\Pbb}{\mathbb P}

\newcommand{\betamin}{\beta_{min}}
\newcommand{\betamax}{\beta_{max}}
\newcommand{\alphamin}{\alpha_{min}}
\newcommand{\alphamax}{\alpha_{max}}

\newtheorem{theorem}{Theorem} 
\newtheorem{proposition}[theorem]{Proposition} 
\newtheorem{remark}[theorem]{Remark} 
\newtheorem{definition}[theorem]{Definition} 
\newtheorem{lemma}[theorem]{Lemma} 

\newcommand{\gammabar}{\overline{\gamma}}
\newcommand{\Rzero}{\mathcal{R}_0}

\title{The impact of recovery rate heterogeneity in achieving herd immunity}
\author{Gabriel Turinici\thanks{CEREMADE, Université Paris Dauphine - PSL, Paris, FRANCE, \textrm{https://turinici.com}, \textrm{Gabriel.Turinici@dauphine.fr}} }
\date{\today}

\begin{document}
	
\maketitle
	
\begin{abstract}
Herd immunity is a critical concept in epidemiology, describing a threshold at which a sufficient proportion of a population is immune, either through infection or vaccination, thereby preventing sustained transmission of a pathogen. In the classic Susceptible-Infectious-Recovered (SIR) model, which has been widely used to study infectious disease dynamics, the achievement of herd immunity depends on key parameters, including the transmission rate ($\beta$) and the recovery rate ($\gamma$), where $\gamma$ represents the inverse of the mean infectious period. While the transmission rate has received substantial attention, recent studies have underscored the significant role of $\gamma$ in determining the timing and sustainability of herd immunity. Additionally, it is becoming increasingly evident that assuming $\gamma$ as a constant parameter might oversimplify the dynamics, as variations in recovery times can reflect diverse biological, social, and healthcare-related factors. 

In this paper, we investigate how heterogeneity in the recovery rate affects herd immunity. We show empirically that the mean of the recovery rate is not a reliable metric for determining the achievement of herd immunity. Furthermore, we provide a theoretical result demonstrating that it is instead the mean 
recovery time, which is the mean
of the inverse $1/\gamma$ of the recovery rate that is critical in deciding whether herd immunity is achievable within the SIR framework.
A similar result is proved for the SEIR model. These insights have significant implications for public health interventions and theoretical modeling of epidemic dynamics.
\end{abstract}

\section{Introduction}
The SIR model, initially developed by Kermack and McKendrick in 1927, remains a cornerstone in mathematical epidemiology, providing a simple yet powerful framework to explore the spread of infectious diseases \cite{kermack1927}. Within this model, individuals transition through three compartments: Susceptible (S), Infectious (I), and Recovered (R), with transitions governed by parameters \(\beta\) (transmission rate) and \(\gamma\) (recovery rate). Herd immunity is traditionally defined in terms of a threshold condition, \(1 - \frac{1}{\Rzero}\), where \(\Rzero = \frac{\beta}{\gamma}\) denotes the basic reproduction number. This threshold suggests that for the epidemic to halt spontaneously, the fraction of immune individuals must exceed \(1 - \frac{1}{\Rzero }\).

The role of \(\gamma\), the recovery rate, in the achievement of herd immunity is often underestimated despite the fact that \(\gamma\) influences not only the pace of recovery but also the duration an individual remains infectious, thereby affecting the overall spread of the disease. A higher \(\gamma\) reduces the average infectious period, consequently lowering the effective reproduction number at any point in time. Conversely, lower values of \(\gamma\) prolong the infectious period, increasing the potential for transmission and delaying the achievement of herd immunity.

In classical models, \(\gamma\) is generally assumed to be constant. However, this simplification does not account for inter-individual variability in recovery times due to differences in biological responses, healthcare access, and treatment protocols. Studies by Anderson and May \cite[chap 8-11 \& 19]{anderson1991}, and Keeling and Rohani \cite[section 6.2.3]{keeling2008} have demonstrated that introducing variability in model parameters, especially in \(\gamma\), can alter the epidemic trajectory and herd immunity threshold. Recent work has extended this approach by treating \(\gamma\) as a random variable, representing the heterogeneity in recovery rates. For instance, Britton et al. (2020) investigated how heterogeneity in both susceptibility and recovery rates affects the level at which herd immunity is achieved \cite{britton2020}.

Other authors also recognized the importance of heterogeneity in epidemic modeling: \cite{arruda_impact_heterogeneous_gamma20} applied heterogeneity primarily in the context of transmission networks, \cite{gou_how_2017}  highlighted its impact  on a structured population and through empirical simulations and \cite{heterog_contact21} recognized its importance in contact rate modeling.

In a series of works, \cite{lloyd_stoch_gamma1_01,lloyd_stoch_gamma2_01}
A.L. Lloyd discussed the heterogeneity in $\gamma$ by introducing several 
intermediary steps into the evolution 
and shows that the stochasticity plays an important role into the propagation dynamics.

Several studies have explored the incorporation of heterogeneity in epidemiological modeling to more accurately reflect real-world disease dynamics. Gourieroux \& Lu (2020) \cite{gourieroux2020sirmodelstochastictransmission} extend the SIR model by introducing nonlinear stochastic transmission, thereby capturing fluctuating reproduction numbers and emphasizing that herd immunity is not a stable outcome in discretized models with stochastic effects. In contrast, Kloeden \& Kozyakin (2011) \cite{kloeden2011dynamics} investigate the impact of non-autonomous, random coefficients, focusing on time-varying dynamics that produce pullback attractors but do not apply stochasticity to recovery rates, as is our focus here. Berestycki et al. (2023) \cite{berestycki_epidemic_2023} examine a reaction–diffusion-based model structured around contact heterogeneity, allowing for variable risk exposure during epidemics but concentrating on contact rates rather than recovery rates.

Almeida et al. (2021) \cite{perthame21} formulate a SEIR model for heterogeneous populations, where recovery rates can vary but are bounded by a minimum threshold, which constrains heterogeneity in a way that differs from our unbounded random-variable approach for $\gamma$. Their analysis is primarily focused on final epidemic size, leaving less attention to comparisons involving recovery heterogeneity. Finally, Novozhilov (2008) \cite{novozhilov_spread_2008} uses heterogeneous transmission parameters in an SIR model, briefly acknowledging stochasticity in recovery rates without developing this aspect further. His work highlights nonlinear transmission dynamics, illustrating how heterogeneity in susceptibility can yield emergent properties that simple moment-closure approaches may overlook.

In this paper, we further explore the impact of heterogeneous recovery rates on herd immunity. We first examine empirically how herd immunity is affected by heterogeneity in the recovery rate and demonstrate that the mean $\gammabar=\Ebb[\gamma]$  of the recovery rate is an inadequate metric for determining herd immunity in this context. 
In particular, we present an example of three cases having identical average recovery rates, $\gammabar$, yet only one exhibits herd immunity; this means  that the average recovery rate alone is insufficient to determine herd immunity.
Subsequently, we provide a theoretical result indicating that the fact that the mean 
recovery time $\Ebb \left[\frac{1}{\gamma(x)} \right]$ 
 being finite is the right criterion in determining whether herd immunity is achievable in the model. This result highlights the importance of considering heterogeneity in epidemic modeling and has significant implications for understanding and managing infectious disease dynamics.

\section{Heterogeneous SIR model}

We follow \cite{novozhilov_spread_2008,perthame21}  and consider a SIR model 
 \cite{kermack1927} in a population that has heterogeneous recovery rate. This means that $\gamma$ is not constant.
 We assume that $\gamma$  results from some individual characteristics and also probably some individual randomness, but we suppose no coupling between the recovery rate and overall societal epidemic situation. 
 This means that, similar to  \cite{perthame21}, 
we can assume $\gamma$ is a deterministic function of the individual (or some ``trait'')  denoted $x$ : $x \in \Omega \mapsto \gamma(x)$.
Introducing as usually $S(t,x)$,  $I(t,x)$ and  $R(t,x)$ 
to be the probability of the individual  $x$ to be 
at time $t$
in the ``Susceptible'',``Infected and infectious''\footnote{To distinguish SIR from the SEIR model used later we will call the middle class ``Infected and infectious'' rather than just ``Infected'' or ``Infectious'' as one may encounter in the literature.} or ``Recovered'' classes respectively, we can write:
\begin{align}
& \partial_t S(t,x) = -S(t,x) \int_\Omega \beta(x,y)I(t,y)dy, & S(0,x)&= S_0(x) 
\label{eq:sir_heterogGamma_S} \\
& \partial_t I(t,x) = S(t,x) \int_\Omega \beta(x,y)I(t,y)dy-\gamma(x)I(t,x), &I(0,x)&= I_0(x) 
\label{eq:sir_heterogGamma_I} \\
& \partial_t R(t,x) = \gamma(x)I(t,x), &R(0,x)&= R_0(x).
\label{eq:sir_heterogGamma_R}
\end{align}
With the notations
\begin{align}
S(t) = \E_x [S(t,x)] =  \int_\Omega S(t,y)dy, 
I(t) = \E_x [I(t,x)], 
R(t) = \E_x [R(t,x)] 
\label{eq:def_siravg}
\end{align}
we can also write
\begin{align}
	&  S'(t) = - \int_\Omega \int_\Omega \beta(x,y) S(t,x) I(t,y)dxdy, & S(0)&= S_0, 
	\label{eq:sir_heterogGamma_ES} \\
	&  I'(t) = \int_\Omega \int_\Omega \beta(x,y) S(t,x) I(t,y)dxdy - \int_\Omega\gamma(y)I(t,y) dy, &I(0)&= I_0 
	\label{eq:sir_heterogGamma_EI} \\
	& R'(t) = \int_\Omega\gamma(y)I(t,y) dy, &R(0)&= R_0.
	\label{eq:sir_heterogGamma_ER}
\end{align}

Note that, in the particular situation when 
$\beta$ and $\gamma$ are constants, 
$\Omega$ can be taken to be a singleton $\{x_0\}$ and 
\eqref{eq:sir_heterogGamma_ES}-\eqref{eq:sir_heterogGamma_ER}
corresponds to the
 standard SIR model. In this case it is known that herd immunity is always present
as soon as $\gamma(x_0)>0$, that is $S(\infty)>0$; the existence of a limit is straightforward because $S(t)$ is decreasing and positive.

The set $\Omega$ designates the ensemble of the ``traits''. It can be thought as being included in some $\R^n$ for $n \ge 1$. In practice these are the parameters required to describe the transmission parameter $\beta$ and recovery  parameter  $\gamma$ which may be heterogeneous within the population. In general a discrete $\Omega$ corresponds to groups of individuals and a continuous $\Omega$ would indicate that there is a continuum of behaviors and characteristics in the population. Mixed situations can also arise if one combines for instance a continuous variable such as age with discrete one (sex, work status and so on) to obtain for instance $\Omega = $[0,100]$ \times \{$``F'',``M''$\} \times ...$.
The function $\beta : \Omega \times \Omega \to \R$ describes on one hand how 
often people susceptible with trait $x$ are likely to meet with infected people with trait $y$ and on the other hand  what are the chances that such a contact is infectious.\footnote{Note that we assume that the contact rates described by $\beta(\cdot,\cdot)$ are 
independent of the epidemic status (this is a hypothesis shared by many contact rate epidemic models).} In practice its values are positive and usually bounded.

For obvious modeling reasons it is standard to suppose $\gamma$ is non-negative~:
\begin{align}
	& \gamma(x)  \ge 0, &\forall x \in \Omega.
	\label{eq:hyp_gamma}
 \end{align}
We will moreover assume the following:
\begin{align}
	&\exists \betamin, \betamax \text{ such that } \forall x \in \Omega : 
	0 <  \betamin \le \beta(x,y)  \le \betamax < \infty,  \ a.e. \  y \in \Omega
	\label{eq:hyp_beta} \\
	& R(0)=0 \label{eq:hyp_RO} \\
	& I(0)>0. \label{eq:hyp_IO}
 \end{align}
Of course, hypothesis \eqref{eq:hyp_IO} is required for something meaningful to arrive, otherwise the dynamic in 
\eqref{eq:sir_heterogGamma_S}-\eqref{eq:sir_heterogGamma_R}
remains constant. Hypothesis  \eqref{eq:hyp_RO} is just a convention of how we count the individuals because persons in the `Recovered' class do not interact any more with anyone else so whoever is already in this class at $t=0$ is not taken into account any more.
Hypothesis 	\eqref{eq:hyp_beta} is mainly needed for technical reasons and also to prevent some unrealistic situations such as totally immune individuals or infinite propagation speed; for this work we wanted to focus more on $\gamma$ and not on pathological $\beta$ cases, 
nevertheless it would be interesting to see how results of the paper change if this technical condition is not satisfied any more.

\subsection{Herd immunity failure: empirical examples} \label{sec:numerical_herd_failure}

Dealing with heterogeneous $\gamma$ is not straightforward and we would like to find a proxy that conveys most of the information contained in the distribution of this parameter from the perspective of herd immunity. To this end we introduce the definition:
\begin{definition}
	We say that \eqref{eq:sir_heterogGamma_S}-\eqref{eq:sir_heterogGamma_R} has the herd immunity property if
	$S(\infty)>0$.
\end{definition}
A naive proposal is the average $\gammabar = \E[\gamma]$. We will show that this is not a good choice. 

\begin{figure}[htb!]
\centering
\includegraphics[width=.32\linewidth]{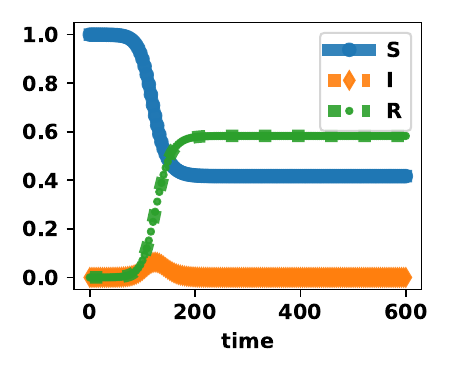}
\includegraphics[width=.32\linewidth]{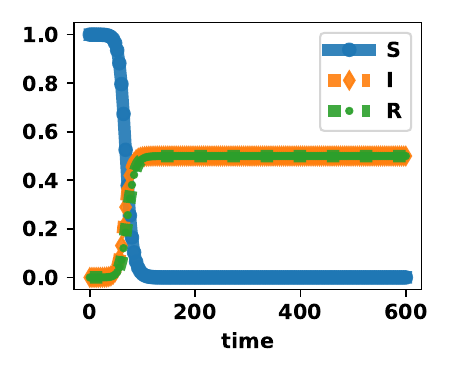}
\includegraphics[width=.32\linewidth]{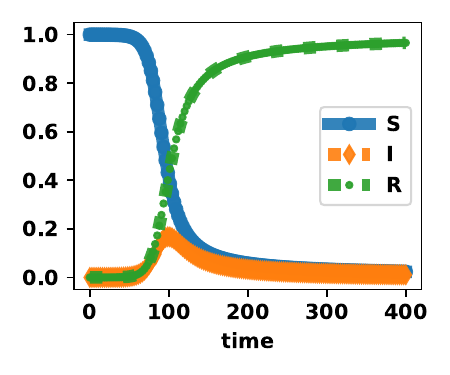}
\caption{{\bf Left:} Solution of the standard SIR model for  $\beta=1/4$, $\gamma=1/6$. Here $S(\infty)\simeq 0.42$.
{\bf Middle:} solution of a two group SIR model for  $\beta=1/4$, $\gamma=0$ or $1/3$ (each with probability $1/2$). Here $S(\infty)= 0.0$.	
{\bf Right:} solution of a SIR model with $\beta=1/4$, $\Omega=[0,1]$, $\gamma(x)=x/3$. Again $S(\infty)= 0.0$.	
}
	\label{fig:sir_hi_failure}
\end{figure}
Consider first the simple example where $\beta=1/4$, $\gamma=1/6$ (both are constants) that corresponds to a 
$\Rzero=3/2$. The numerical result is given in figure \ref{fig:sir_hi_failure} (left) where
it is seen that almost one half of the initial population does not become sick even in absence of any containment measures. This means that herd immunity is active. 
Consider now the almost similar situation of a two group SIR model 
where half of the people have $\gamma=0$ and half of them have $\gamma=1/3$. As a random variable $\gamma$ has the same mean as before, but the evolution is completely different
 (see middle plot of figure \ref{fig:sir_hi_failure}).
In this case the susceptible class is emptied completely i.e., $S(\infty)=0$ therefore the herd immunity is not present any more. We formalize this is the following result:
\begin{lemma}
Consider $\Omega=\{1,2\}$ with values $1$ and $2$ having probability $1/2$ each
and assume that $\gamma(1)=0$  and  $\gamma(2)>0$. Then  $S(\infty)=0$.
\end{lemma}
\begin{proof}
In this situation the SIR system (written only for the first two classes) is 
\begin{align}
& S'(t,g) = - \beta S(t,g) I(t), \ g=1,2 
\\ & 
I'(t,g) = \beta S(t,g) I(t) - \gamma(g) I(t,g)  \ g=1,2 
\\ & 
\text{ where } S(t)= \frac{S(t,1)+S(t,2)}{2}, \ I(t)= \frac{I(t,1)+S(t,2)}{2}.
\end{align}
Since $\gamma(1)=0$ we have that for all $t\ge 0$:
 $I'(t,1)= \beta S(t,1) I(t) \ge 0$ so $I(t,1)$ is increasing. Moreover 
since $I(0) >0$ we have $I(0,1)>0$; all this means that there exist some $\varsigma$ such that  $I(t,1)>\varsigma$ for all $t\ge t_\varsigma$.
On the other hand for $t\ge t_\varsigma$:  $S'(t) = -\beta S(t) I(t) \le  -\beta S(t) I(t,1) \le -\beta S(t) \varsigma$ which 
allows to write $S(t) \le S(t_\varsigma) e^{-\sigma (t-t_\varsigma)}$
and shows that $S(\infty)=0$.
\end{proof}

For this particular case one could imagine that some computation involving the $\Rzero$ of each group could be used as a warning, because for the group with $\gamma=0$ its 
$\beta/\gamma=\infty$. In fact this is not the correct definition of $\Rzero$ and moreover other situations exist where $\gamma$ is (almost everywhere) strictly positive but still  herd immunity is failing. 
To see that, consider $\Omega=[0,1]$, $\gamma(x)=x/3$. 
The average of $\gamma$ remains $1/6$. The solution is displayed 
in the right plot of figure	\ref{fig:sir_hi_failure} and we see that again $S(\infty)=0$.

As a side remark, we see that in situation 2 (middle plot of Figure~\ref{fig:sir_hi_failure}) the Infected class remains high at infinity and in situation 3 
(right plot of Figure~\ref{fig:sir_hi_failure}) 
the infected class goes to zero. We will see in Propositions~\ref{prop:SIR_herd_immunity} and 
\ref{prop:SEIR_herd_immunity} that
this behavior is generic. 
\bigskip

\noindent {\bf Details of the numerical implementation:} the Python program used to obtain the results of the paper is available at \\ 
{
https://github.com/gabriel-turinici/epidemiology\_heterogeneous\_gamma}

\section{Herd immunity and finite average recovery time}

We are now ready to give an important result of this work stating that the herd immunity is equivalent to  $1/\gamma$ having a finite average. 

\begin{proposition} 
Assume	\eqref{eq:hyp_gamma}, \eqref{eq:hyp_beta}, \eqref{eq:hyp_RO} and	\eqref{eq:hyp_IO} are satisfied. Then 	\eqref{eq:sir_heterogGamma_S}-\eqref{eq:sir_heterogGamma_R}
has the herd immunity property if and only if~:
\begin{align}
& \Ebb \left[\frac{1}{\gamma(x)} \right] < \infty \hfill 
\text{ (finite mean recovery time)}.
\label{eq:finite_inverse_gamma}
\end{align}
\label{prop:SIR_herd_immunity}
\end{proposition}
\begin{remark}
Since $1/\gamma$ can be interpreted as mean recovery time for an individual, the 
relation
\eqref{eq:finite_inverse_gamma} is nothing else than requiring that the average recovery time across the population be finite.  In particular we will prove that if a non-negligible population sub-group has infinite recovery time 
 (one example is the middle plot of Figure~\ref{fig:sir_hi_failure})
then all population will be infected (no herd immunity). 
But the conclusion is more general as we can have 
$\Ebb \left[\frac{1}{\gamma(x)} \right] = \infty$ with $\gamma(x)$ being finite for any $x$, as seen in Section~\ref{sec:numerical_herd_failure}, 
right plot of Figure~\ref{fig:sir_hi_failure}. 
\end{remark}
	
\begin{proof} 
 We will denote
\begin{align}
	Z = \left\{ x : \gamma(x)=0 \right\},
	\label{eq:def_zero_gamma_set}
\end{align}	
and note that  because of the hypothesis \eqref{eq:hyp_gamma} the term $ \E \left[\frac{1}{\gamma(x)} \right] $
in \eqref{eq:finite_inverse_gamma} is well defined (but not necessarily finite) with the convention that $1/\gamma(x) = +\infty$ whenever $x \in Z$.
\\
{\bf Part I: \eqref{eq:finite_inverse_gamma} implies herd immunity}
\\ 
Suppose $\Ebb \left[\frac{1}{\gamma(x)} \right] < \infty$. In particular $\Pbb \left[Z \right]=0$ so we can neglect all such $x$.
Equation \eqref{eq:sir_heterogGamma_S} 
can be written 
\begin{align}
& \frac{d}{dt}\ln(S(t,x)) = -\int_\Omega \beta(x,y) I(t,y) dy =-\int_\Omega \frac{\beta(x,y)}{\gamma(y)} \frac{d}{dt} R(t,y) dy
\label{eq:dlnS}
\end{align}
which means that 
\begin{align}
& S(t,x)= S(0,x) \cdot exp\left\{
-\int_0^t\int_\Omega \frac{\beta(x,y)}{\gamma(y)} \frac{d}{ds} R(s,y) dy ds	\right\} 
\nonumber \\&
=	S(0,x) \cdot exp\left\{\int_\Omega \frac{\beta(x,y)}{\gamma(y)}  (R(0,y)-R(t,y)) dy 	\right\} 
\label{eq:stx1} \\&
\ge  	S(0,x) \cdot exp\left\{- \int_\Omega \frac{\betamax}{\gamma(y)}  dy 	\right\}, \label{eq:stx2}
\end{align}
where we used the hypothesis \eqref{eq:hyp_beta}, \eqref{eq:hyp_RO} 
and $R(t,y)\in[0,1]$ (it is a probability).
By taking averages over $x$ we obtain 
\begin{align}
	& S(\infty) \ge S(0) \cdot exp\left\{- \int_\Omega \frac{\betamax}{\gamma(y)}  dy 	\right\} > 0, 
\end{align}
since  $\Ebb \left[\frac{1}{\gamma(x)} \right] < \infty$. This proves the herd immunity.
\\
{\bf Part II: herd immunity implies \eqref{eq:finite_inverse_gamma}} 

\noindent Let us 
first study the situation when $\gamma(x)=0$ on a non-negligible set, or equivalently, 
$\Pbb \left[ Z  \right] >0$. 
Note that $I(t)$ is a continuous function so, since $I(0)>0$ it will still be strictly positive at least on some interval $[0,\epsilon]$. For  $x\in Z$ the equation \eqref{eq:sir_heterogGamma_I}
becomes
$\partial_t I(t,x) = S(t,x) \int_\Omega \beta(x,y)I(t,y)dy$ while 
\eqref{eq:sir_heterogGamma_R}
reduces to 
$\partial_t R(t,x) = 0$. Thus, for such $x$, for any $t\ge 0$: $R(t,x) = 0$; since we also have 
$S(t,x)+ I(t,x)+R(t,x)=1$ 
we conclude that  
\begin{align}
& \partial_t I(t,x) = (1-I(t,x))\int_\Omega \beta(x,y)I(t,y)dy
\nonumber 
\ge (1-I(t,x))\betamin I(t)
\end{align}
and therefore, denoting $I_Z(t) = \int_Z I(t,y) dy$ (recall definition \eqref{eq:def_zero_gamma_set}) we obtain
\begin{align}
	& \frac{d}{dt} I_Z(t) 	\ge (1-I_Z(t))\betamin I(t)
	\ge \betamin \cdot \Pbb(Z) \cdot (1-I_Z(t))  I_Z(t) ,	
\end{align}
which shows that $I_Z(t) \to 1$ as $t\to \infty$ and in particular for some $\epsilon >0$ :  $I_Z(t) \ge \epsilon$ as soon as $t\ge t_\epsilon$.
But since, for a general $x\in \Omega$: 
\begin{align}
	& \frac{d}{dt}\ln(S(t,x)) = -\int_\Omega \beta(x,y) I(t,y) dy 
 \le  -\betamin  \cdot \Pbb(Z) \cdot \epsilon  
\end{align}
we obtain $S(t_2,x) \le S(t_1,x)\cdot exp( -(t_2-t_1) \cdot \betamin  \cdot \Pbb(Z) \cdot \epsilon )$ for any $t_1,t_2 \ge t_\epsilon$ for any $x\in \Omega$. By integration over $x$ this also means
 $S(t_2) \le S(t_1)\cdot exp( -(t_2-t_1) \cdot \betamin  \cdot \Pbb(Z) \cdot \epsilon )$. When $t_2 \to \infty$ we obtain $S(\infty)=0$ in contradiction with the hypothesis. Therefore $\Pbb(Z)$ must be null. So in the following we can discard all $x\in Z$ as they contribute negligible to any integration over $\Omega$. 
\\
Since $I(0)> 0$ it will remain larger than $ I(0)/2$ on some
interval $[0,\epsilon]$; 
from \eqref{eq:dlnS} we can write:
\begin{align}
&
 \forall t \le \epsilon~: 
 \frac{d}{dt}\ln(S(t,x)) = -\int_\Omega \beta(x,y) I(t,y) dy \le -\betamin I(t) \le -\betamin \frac{I(0)}{2},
 \label{eq:proof_sir_part22_1}
\end{align}
which means that
for some constant $C_3$ independent of $x$ with $C_3< 1$ we have  $S(\infty,x) \le S(\epsilon,x)\le C_3$. On the other hand it can be easily checked that $\lim_{t\to \infty} I(t,x)=0$ (convergence may be non-uniform in $x$). Using the two relations we obtain that 
 $R(\infty,x)\ge 1-C_3$.
We use now again~\eqref{eq:stx1} and recall that $R(0,y)=0$ for all $y$ (hypothesis \eqref{eq:hyp_RO}) to obtain~:
\begin{align}
	& S(\infty,x)= 	S(0,x) \cdot exp\left\{\int_\Omega \frac{\beta(x,y)}{\gamma(y)}  (R(0,y)-R(\infty,y)) dy 	\right\} 
	\label{eq:stx1-v2} \\&
	\le  	S(0,x) \cdot exp\left\{- \int_\Omega \frac{\betamin}{\gamma(y)} C_3 dy 	\right\}.
 \label{eq:proof_sir_part22_3}
\end{align}
Taking average over $x$ and using $S(\infty)>0$ we obtain that 
$\Ebb \left[\frac{1}{\gamma(x)} \right]$ must be finite which is the conclusion.
\end{proof}

\begin{figure}[htb!]
	\centering
	\includegraphics[width=.32\linewidth]{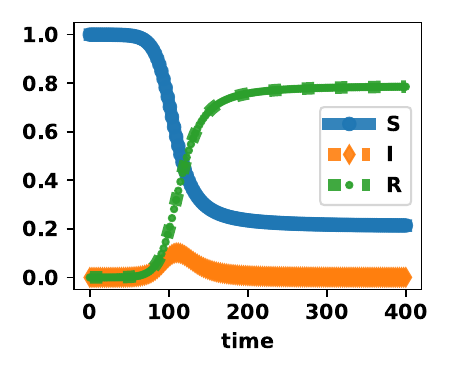}
\caption{Solution of a SIR model with $\beta=1/4$, $\Omega=[0,1]$, $\gamma(x)= \frac{3}{2} \sqrt{x}$. Herd immunity ($S(\infty)>0$) is observed numerically, 
coherent with the theoretical results.}
\label{fig:sir_herd_ok}
\end{figure}

As a numerical application we revisit a situation similar to the one in Section~\ref{sec:numerical_herd_failure} but here we choose $\gamma(x)= \frac{3}{2} \sqrt{x}$. This has the same average $\gammabar = \E[\gamma]=1/6$ as above but now
$\Ebb \left[\frac{1}{\gamma(x)} \right] = 4/3 < \infty$. Proposition~\ref{prop:SIR_herd_immunity}  
states that herd immunity must be present, which is confirmed by 
the numerical results in Figure~\ref{fig:sir_herd_ok}.

\subsection{Herd immunity : extension to the SEIR model}
Similar results can be proved for the SEIR model. Recall that in this case the ``Infected and infectious'' class is split in two parts, the
``Exposed'' class that is infected but not yet infectious and the 
 ``Infectious" class that is infected and infectious. The equations analogue to 	\eqref{eq:sir_heterogGamma_S}-\eqref{eq:sir_heterogGamma_R} read now:
\begin{align}
	& \partial_t S(t,x) = -S(t,x) \int_\Omega \beta(x,y)I(t,y)dy, & S(0,x)&= S_0(x) 
	\label{eq:seir_heterogGamma_S} \\
	& \partial_t E(t,x) = S(t,x) \int_\Omega \beta(x,y)I(t,y)dy-\alpha(x)E(t,x), &E(0,x)&= E_0(x) 
	\label{eq:seir_heterogGamma_E} \\
	& \partial_t I(t,x) = \alpha(x)E(t,x)-\gamma(x)I(t,x), &I(0,x)&= I_0(x) \label{eq:seir_heterogGamma_I} \\
	& \partial_t R(t,x) = \gamma(x)I(t,x), &R(0,x)&= R_0(x).
	\label{eq:seir_heterogGamma_R}
\end{align}
Similar equations hold for the aggregate quantities $S(t)$, $E(t)$, $I(t)$, $R(t)$. We add two hypothesis which are natural additions or replacements of the previous ones:
\begin{align}
	&\exists \alphamin, \alphamax \text{ such that: }  
	0 <  \alphamin \le \alpha(x)  \le \alphamax < \infty,  \forall x \in \Omega
	\label{eq:hyp_alpha} \\
	& E(0)+I(0)>0. \label{eq:hyp_E0IO}
\end{align}
\begin{proposition} 
	Assume	\eqref{eq:hyp_gamma}, \eqref{eq:hyp_beta}, \eqref{eq:hyp_RO},  
	\eqref{eq:hyp_alpha},
	 and	\eqref{eq:hyp_E0IO} are satisfied. Then 
	the SEIR system	\eqref{eq:seir_heterogGamma_S}-\eqref{eq:seir_heterogGamma_R}
	has the herd immunity property if and only if \eqref{eq:finite_inverse_gamma} holds.
\label{prop:SEIR_herd_immunity}
\end{proposition}
\begin{proof}
We use the same notation for $Z$ as in \eqref{eq:def_zero_gamma_set} and more generally follow, when possible, the proof of Proposition~\ref{prop:SIR_herd_immunity}.

{\bf Part I: \eqref{eq:finite_inverse_gamma} implies herd immunity}
\\ 
With similar computations and estimations as in \eqref{eq:dlnS}, 
\eqref{eq:stx1} and 
\eqref{eq:stx2} we can write 
\begin{align}
& S(t,x)	\ge  	S(0,x) \cdot exp\left\{- \int_\Omega \frac{\betamax}{\gamma(y)}  dy 	\right\}, 
\end{align}
and furthermore  because 
$\Ebb \left[\frac{1}{\gamma(x)} \right] < \infty$:
\begin{align}
	& S(\infty) \ge S(0) \cdot exp\left\{- \int_\Omega \frac{\betamax}{\gamma(y)}  dy 	\right\} > 0, 
\end{align}
which proves the herd immunity.
\\
{\bf Part II: herd immunity implies \eqref{eq:finite_inverse_gamma}} 

\noindent We consider first the situation $\Pbb \left[ Z  \right] >0$. The difference with respect to the SIR proof is to consider, 
for $x\in Z$, the evolution of 
 $E(t,x)+I(t,x)$ instead of $I(t,x)$: 
\begin{align}
	& \partial_t (E+I)(t,x) = (1-(E+I)(t,x))\int_\Omega \beta(x,y)I(t,y)dy
	\nonumber \\& 
	\ge (1-(E+I)(t,x))\betamin I(t)
	\ge \betamin  \cdot P(Z) \cdot (1-(E+I)(t,x))I_Z(t).
\end{align}
By summing over $Z$ and with notation 
$E_Z= \int_Z E(t,x) dx$ (and  $I_Z$ as before)
we conclude that:
\begin{align}
	& \frac{d}{dt}(E_Z+I_Z)(t) \ge \betamin  \cdot P(Z) \cdot (1-(E_Z+I_Z)(t))I_Z(t).
\end{align}
Since on the other hand we also obtain directly from \eqref{eq:seir_heterogGamma_I}
$ \frac{d}{dt}I_Z(t) \ge \alphamin E_Z(t)$
we conclude that $E_Z(\infty)=0$, $I_Z(\infty)=1$ and as before a contradiction. This means that $\Pbb(Z)>0$ is not compatible with herd immunity. 

When $\Pbb[Z]=0$ the conclusion follows by  repeating the arguments in equations  \eqref{eq:proof_sir_part22_1}-\eqref{eq:proof_sir_part22_3}.
\end{proof}

\section{Discussion and conclusion}

In this study, we have explored the critical role of heterogeneity in recovery rates within the context of herd immunity for the traditional SIR and SEIR models. Our findings challenge the naive intuition that the {\bf mean recovery rate} $\gammabar$ alone is sufficient for determining the conditions under which herd immunity can be achieved. Instead, we have shown, under appropriate hypotheses, that the {\bf mean recovery time} $\mathbb{E} \left[\frac{1}{\gamma} \right]$  
serves as the key criterion for determining the attainability of herd immunity:  herd immunity is achievable if and only if $\mathbb{E} \left[\frac{1}{\gamma} \right]$ 
is finite.

These results highlight the importance of incorporating realistic variations in recovery times into epidemiological models, as such heterogeneity can significantly influence disease dynamics and the sustainability of herd immunity. 
%

Future work could further refine these findings by examining how other forms of heterogeneity, such as age-specific recovery patterns or differing healthcare access interact with vaccination strategies and natural immunity. Some of our technical hypothesis may also deserve future examination.
By embracing these complexities, we can enhance our understanding of epidemic propagation and design more effective interventions to mitigate the spread of infectious diseases.

\bigskip

\noindent {\bf Disclosure : } The author declares  no  
conflict of interest related to the results present in this work.

\noindent {\bf Data availability} No proprietary dataset was used in this paper. The implementation link provided above contains all information required to replicate the results.

\end{document}